# Homonuclear J-Coupling Spectroscopy using J-Synchronized Echo Detection

Stephen J. DeVience[a,*] and Matthew S. Rosen[b,c]

a. Scalar Magnetics, LLC, 3 Harolwood Ct., Apt C, Windsor Mill, MD 21244, USA.

b. Athinoula A. Martinos Center for Biomedical Engineering, Massachusetts General Hospital, 149th Thirteenth St., Charlestown, MA 02129, USA

c. Department of Physics, Harvard University, 17 Oxford St., Cambridge, MA 02138, USA.

*Corresponding author. Address: Scalar Magnetics, LLC, 3 Harolwood Ct., Apt C, Windsor Mill, MD 21244, USA. stephen@scalarmag.com

Email addresses: stephen@scalarmag.com (S. J. DeVience), msrosen@mgh.harvard.edu (M. S. Rosen)

**Keywords**: J-Coupling; CPMG; spin echo train; spin-lock induced crossing

## Abstract

In the strong coupling regime with J-coupling much larger than chemical shift differences, J-coupling spectroscopy enables spectral identification of molecules even when conventional NMR fails. While this classically required the presence of a heteronucleus, we recently showed that J-coupling spectra can be acquired in many homonuclear systems using spin-lock induced crossing (SLIC). Here, we present an alternative method using a spin echo train in lieu of a spin-locking SLIC pulse, which has a number of advantages. In particular, spin echo acquisition within the pulse train enables simultaneous collection of time and frequency data. The resulting 2D spectrum can be used to study dynamic spin evolution, and the time domain data can be averaged to create a 1D J-coupling spectrum with increased signal-to-noise ratio.

## Introduction

To create easily interpreted spectra, NMR spectroscopy is typically performed under weak coupling conditions where the chemical shift frequency differences $\Delta \nu$ are significantly larger than the scalar couplings, $J$. However, magnetic fields >1 T are typically required to create sufficient frequency dispersion. The ability to acquire high-resolution spectra at lower magnetic fields would be advantageous in many situations, such as for benchtop and educational instruments [1-3], portable operations for oil-field exploration [4-5], spectroscopy in the presence of ferromagnetic and paramagnetic substances [6], optically-detected NMR with nitrogen vacancies as sensors [7-8], and chip-scale spectrometers [9-11]. Unfortunately, as field strength is reduced spectral information becomes difficult to interpret and is eventually lost, because in the strong-coupling regime (when $J \gg \Delta v$) the NMR spectrum of any homonuclear spin system exhibits only a single dominant spectral line. Although adding a heteronucleus such as $^{13}C$ or $^{31}P$ can enable J-coupling spectroscopy in this regime, it is not a practical solution in most situations, and the signal from natural-abundance isotopomers is small [12-16].

We previously showed that homonuclear J-coupling spectroscopy at low field strengths can be implemented using the spin-lock induced crossing (SLIC) sequence [17-21]. SLIC detects the level anti-crossings created whenever there is at least one chemical shift difference among the spins. In SLIC, very weak spin-locking is applied to the system, followed by acquisition of the free-induction decay (FID) signal. Measurements are repeated for a series of spin-lock amplitudes to create a J-coupling spectrum by plotting FID intensity as a function of spin-lock nutation frequency. Dips in FID intensity indicate the positions of level anti-crossings, which in turn are a reflection of the J-couplings within the molecule. Unfortunately, this procedure is slow and suffers from drawbacks such as sensitivity to $B_0$ and $B_1$ offsets and instability. It is also difficult to extend the technique to two dimensions, for example to simultaneously measure the dependence on spin-locking time and nutation frequency, since the acquisitions must be completely repeated along the time dimension.

In this work, we introduce an alternative method for homonuclear J-coupling spectroscopy based on a train of spin echoes rather than spin-locking. Since the energy levels are detected whenever the pulse timing is J-synchronized, rather than via level anti-crossings, we follow the nomenclature of Pileio, *et al.* and refer to this technique as synchronized echo (abbreviated SyncE in this work). It has a number of significant advantages, most importantly the ability to perform echo acquisitions within the sequence, thereby producing a 2D spectrum in the same amount of time as a typical 1D spectrum with SLIC. The echo acquisitions can be averaged to produce more than a 30-fold improvement in the contrast-to-noise ratio compared to the original

SLIC sequence. Importantly, using either SLIC or synchronized echoes detects the same spin energy levels and produces equivalent J-coupling spectra.

**Theoretical Description**

The theory behind SLIC has been described previously by DeVience, *et al*. and others [18-24]. To summarize, magnetization is first rotated onto the x-axis using a 90° excitation pulse (Fig. 1a). In the rotating frame, the energy levels in the strong-coupling regime consist of sets of degenerate states separated by gaps of the order *J* with eigenstates defined by their total spin quantum number *F*. Spin-locking is then applied, which splits these degenerate states proportionally to their magnetic quantum number $m_F$. At particular nutation frequencies, $v_n$, some of these states exhibit level anti-crossings, and x-axis magnetization ($M_x$) coherently evolves into dark quantum states that do not contribute to the FID signal (Fig. 2a). These crossings are detected as dips in the FID intensity. The frequency and intensity of the dips reflect J-couplings and chemical shift differences in the system, respectively.

The Fourier transform for a train of appropriately spaced pulses is similar to that of CW spin-locking, albeit with additional harmonic components, so one might suppose the response of the spin system to the pulse train should be similar, if not identical. It would therefore be possible to substitute a spin echo train for spin-locking in the SLIC sequence and obtain the same J-coupling spectrum, for example using the sequence shown in Fig. 1b. Analysis of the quantum mechanics shows that this is only partially true, and that in fact synchronized echoes create an equivalent J-coupling spectrum but with dips of twice the intensity.

To understand why, we analyze the sequence following the method of Pileio, *et al*. and Tayler, *et al*., but generalizing to systems with more than two spins [17,25]. In their M2S/S2M pulse sequence for converting magnetization to singlet order in a pair of spin-1/2 nuclei, they used a series of two spin echo trains. The first stage of M2S is a J-synchronized echo train that converts x-axis magnetization into a singlet-triplet coherence and is equivalent to that in Fig. 1b. The sequence begins with a 90° pulse that moves longitudinal magnetization into transverse magnetization along the x-axis ($M_x$). In terms of a density matrix, $M_x$ is a coherence composed of all combinations of matrices with a +/- 1 difference in magnetic spin quantum numbers, i.e. $M_x = \sum_i c_i^2 |F, m_F\rangle\langle F, m_F \pm 1|$. In the case of a singlet state (*F*=0), there is only a single state with $m_F$=0, and so $M_x$ cannot contain singlet terms.

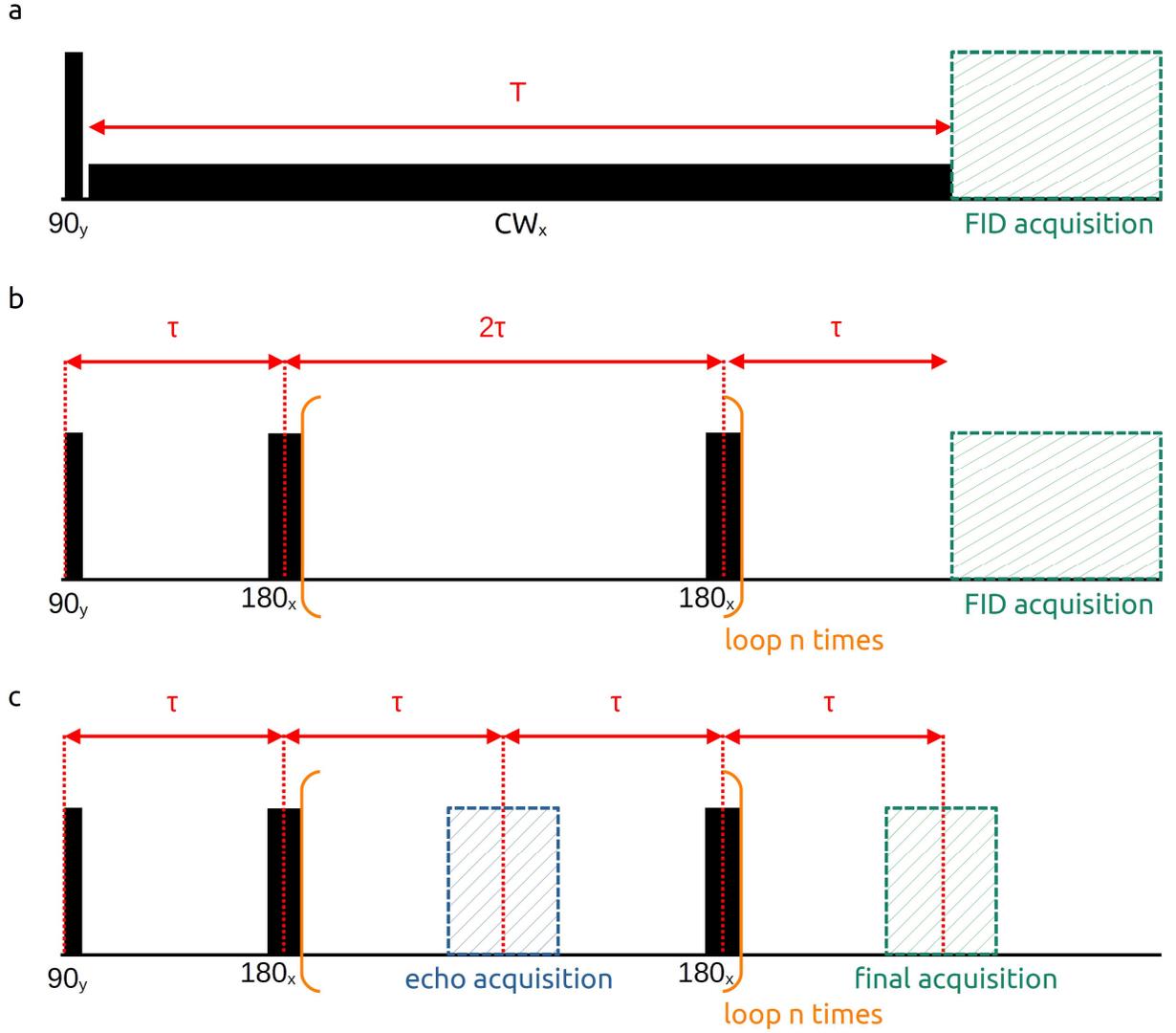

Figure 1: (a) SLIC: Spin-locking is applied for the total time T, followed by FID acquisition. (b) Synchronized echo (SyncE) with a single acquisition: For a series of $\tau$ values, a train of refocusing pulses is played out followed by an FID acquisition. Delay $\tau$ and number of loops $n$ are chosen so that the total time for the echo train, $T = 2\tau(n + 1)$, is always the same. (c) Multi-acquisition synchronized echo: Measurements of echo intensity are acquired within the pulse train, followed by an optional acquisition of the final echo and/or FID.

For a given pair of states, we can create a Bloch sphere and analyze whether the combination of Hamiltonian and pulse sequence can move the state around the Bloch sphere. We find that the pulse train interconverts any two states separated by the selection rule $\Delta F = \pm 1, \Delta m_F = 0$. The simplest example following this rule is the transition between $|T_0\rangle$ and $|S_0\rangle$, or in more explicit form $|1,0\rangle$ and $|0,0\rangle$. These are connected by the interaction term $\langle T_0 | \nu_1 \hat{I}_{1z} - \nu_2 \hat{I}_{2z} | S_0 \rangle = \delta\nu$, where $\delta\nu = \nu_1 - \nu_2$, the difference in resonance frequencies between the two spins. In a Bloch

sphere defined by the states $|T_0\rangle$, $|S_0\rangle$, the interaction term driven by $\delta v$ attempts to move the state around the sphere from $|T_0\rangle$ to $|S_0\rangle$. However, at the same time the quantum state precesses about the longitudinal axis at the frequency gap between these states, $\Delta E = J$. Because $J$ is significantly larger than $\delta v$, only a small amount of $|T_0\rangle$ makes it to $|S_0\rangle$ before precession reverses the process and it returns to $|T_0\rangle$. The state thus makes a small circle near the top of the sphere.

To enable full interconversion between these states, either the energy gap can be removed (as in SLIC), or the effect of precession can be eliminated (using a J-synchronized echo train). In the latter, the spacing between refocusing pulses is chosen so that after a half-cycle of precession the state is brought to the opposite side of the sphere, from where the interaction term can continue the evolution from $|T_0\rangle$ to $|S_0\rangle$. The state is thus kept on the side of the sphere in which evolution proceeds in the correct direction (Fig. 2b).

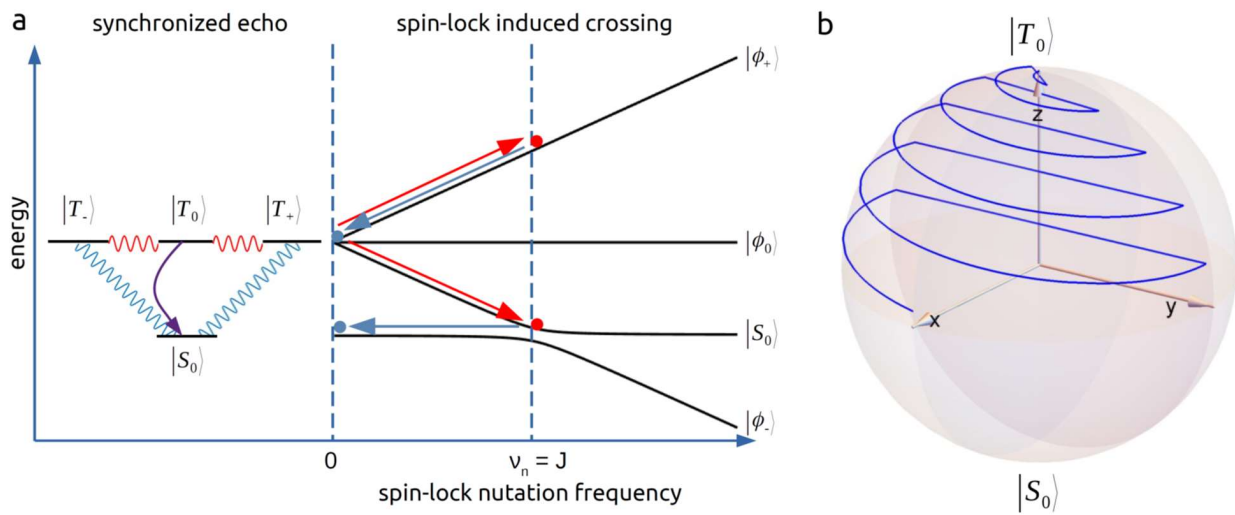

Figure 2: (a) The synchronized echo (SyncE) sequence converts $M_x$, a coherence among triplet states (red), into a singlet-triplet coherence (blue), which is NMR silent. $M_x$ can be completely depleted as it is fully transformed to the singlet-triplet coherence. On the other hand, in SLIC $M_x$ is represented as a population of two spin-locked triplet states (red circles). During spin-locking, the population of one of these states is converted to singlet, so at the end of the sequence 50% of $M_x$ is depleted (blue circles). The other half of $M_x$ can also be depleted via a second block of spin-locking with the opposite phase. (b) Spin echo converts the $|T_0\rangle$ component to $|S_0\rangle$ in a series of stages around the Bloch sphere.

After a full evolution from $|T_0\rangle$ to $|S_0\rangle$, the density matrix has the form

$\sum_i c_i^2 |F \pm 1, m_F\rangle\langle F, m_F \pm 1|$, in this example consisting of coherences between $|S_0\rangle$ and $\langle T_\pm|$, plus their complex conjugate. At this point, the M2S sequence continues with a second pulse train to convert the coherence to a singlet-triplet population difference. We instead perform an FID acquisition to measure the remaining x-axis magnetization $M_x$. The singlet-triplet coherence terms do not appear in $M_x$, so at this point $M_x$ has been depopulated and the spectral line has a lower intensity as a result.

For more complex systems, this same analysis can be performed between every combination of states separated by $\Delta F = \pm 1, \Delta m_F = 0$. For example, in a three spin-1/2 system, the combinations would be: $\left|\frac{3}{2}, \frac{1}{2}\right\rangle$ with $\left|\frac{1}{2}, \frac{1}{2}\right\rangle$ and $\left|\frac{3}{2}, -\frac{1}{2}\right\rangle$ with $\left|\frac{1}{2}, -\frac{1}{2}\right\rangle$. For these transition to be detected, the resulting coherences do not need to contain singlet states or be long-lived in order to be dark states, they simply need to be missing from the definition of $M_x$.

Two differences from SLIC are apparent. First, synchronized echo results in the creation of coherences rather than population differences, and these may have different relaxation properties resulting in different dynamics. Second, synchronized echo creates coherences between $|S_0\rangle$ and both $\langle T_+|$ and $\langle T_-|$, whereas with SLIC the singlet interacts with only one of the three triplet states at a time. Therefore, twice as much magnetization is removed from $M_x$ using the synchronized echo sequence.

**Multi-acquisition Scheme**

In the most straightforward replacement of spin-locking with a pulse train, the remaining x-axis magnetization would be detected following the full evolution period. However, the synchronized echo technique has the additional benefit of gaps between the 180° pulses in which echo acquisitions can be collected (Fig. 1c). By measuring each echo, we can track the dynamics at each step in the evolution around the Bloch sphere while magnetization transfer is still occurring. This has two benefits. First, by measuring $M_x$ over time, the rate of magnetization transfer can be measured and possibly used to determine chemical shift differences and relaxation rates. In contrast, to perform the same type of measurement with SLIC would require repeated experiments with different evolution times. For 2D experiments, echo acquisition therefore greatly decreases the total experiment time. Second, the echo acquisitions can be averaged over time to create a 1D projection of the J-coupling spectrum. Averaging significantly reduces the noise compared with taking a single measurement at the end of the sequence. Moreover, the specific time points to average can be windowed to choose the periods with strongest signals, thereby optimizing the contrast-to-noise ratio.

To create a J-coupling spectrum with synchronized echoes, data are acquired with sequences shown in Fig. 1b or 1c for a series of delay times $\tau$. The equivalent nutation frequency for each measurement is $v_n = 1/(4\tau)$. The number of loops $n$ is chosen so that the total sequence time $T = 2\tau(n+1)$ is constant. Using a constant time $T$ ensures that $T_2$ relaxation effects remain relatively constant throughout the experiment resulting in a smooth background for the J-coupling spectrum. $T_2$ is shorter for larger values of $\tau$, leading to smaller signals for low nutation frequencies, but the resulting background is easily removed. A constant $T$ also ensures a constant linewidth for the dips, as the two are related through a Fourier relationship in the same way as for SLIC. The resolution improves with experiment time and is given by

$$\Delta v_n = \frac{1}{4T/2n+2} - \frac{1}{4T/2n} = \frac{1}{2T} \quad (1)$$

The drawback of using a constant $T$ is that nutation frequency points (and the associated frequency resolution) cannot be chosen arbitrarily as they can be for SLIC.

**Results and Discussion**

To compare techniques, J-coupling spectra of neat ethyl acetate were first measured with SLIC and synchronized echo (SyncE) sequences shown in Figures 1a and 1b respectively. The SLIC pulse and total echo train length was $T = 1$ second, and the FID was acquired following the complete series of pulses. Results are shown in Fig. 3a and 3b along with spectra predicted via simulation. The dips for SyncE were about twice as deep as for SLIC, as predicted by theory (blue curves), while the noise level for SyncE was about 1/3 of the level for SLIC, resulting in a contrast-to-noise ratio (CNR) about 6 times higher using SyncE (Table 1). However, a higher resolution (0.3 vs. 0.5 Hz) was possible with SLIC because the SyncE resolution was constrained by the 1 s sequence duration.

We then implemented the multi-acquisition SyncE scheme (Fig. 1c), collecting data during each echo. We extended the pulse train to 3 seconds to provide higher resolution (0.167 Hz). Figure 3c shows results when the echo data are averaged to create a 1D J-coupling spectrum. The dips are slightly deeper than using the 1 second sequence, and the single-scan noise is about 3 times smaller, giving another factor-of-4 improvement in CNR. This results in a 20 to 30-fold improvement compared with the original SLIC sequence. A spectrum was also acquired using composite 180° pulses (dashed red spectrum), which increased both depth and noise slightly, providing the same CNR results as simple 180° hard pulses.

Table 1. Comparison of contrast-to-noise (CNR) ratios for the two dips of the ethyl acetate J-coupling spectrum acquired with SLIC and synchronized echo (SyncE).

| Sequence | Evolution Time ($T$) | Scans | Single Scan CNR 10.7 Hz | Single Scan CNR 17.7 Hz |
| --- | --- | --- | --- | --- |
| SLIC | 1 s | 8 | 3.5 | 2.2 |
| SyncE FID | 1 s | 8 | 27 | 13 |
| SyncE multi-acquisition | 3 s | 2 | 109 | 52 |
| SyncE multi-acq composite pulses | 3 s | 2 | 110 | 52 |

All spectra acquired with synchronized echoes clearly exhibit harmonic dips near 3.5 Hz and 6 Hz as predicted by the simulations. The Fourier transform of the SLIC pulse contains only one frequency component, making it sensitive to resonances at only the nutation frequency. The transform of a spin echo train contains odd harmonics of decreasing amplitude. As a result, measurement of $\nu_n = 3.5$ Hz is also sensitive to 10.5 Hz, near the frequency of the first ethyl acetate resonance, as well as 16.5 Hz, 14.5 Hz, and so on.

Figure 3d shows the full 2D spectrum of ethyl acetate acquired with the multi-acquisition SyncE scheme using simple 180° rectangular hard pulses. Unlike time domain measurements using SLIC, which showed the predicted coherent oscillations in dip intensity over time [17], measurements with SyncE exhibited a monotonic decrease in intensity. This likely reflects different lifetimes for the dark states, since SLIC moves magnetization into population states while SyncE instead creates coherences. Fast relaxation of the coherences would prevent them from being converted back to $M_x$, as required to produce an oscillation.

To ensure synchronized echo also measures spectra correctly in the presence of chemical exchange, we acquired spectra of ethanol under anhydrous and hydrated conditions using the multi-acquisition scheme. As shown in Figure 4, measured spectra matched well with simulations assuming a 5.0 Hz coupling with the hydroxyl proton in anhydrous ethanol and 0 Hz coupling in a heavily hydrated form (46% ethanol and 54% water).

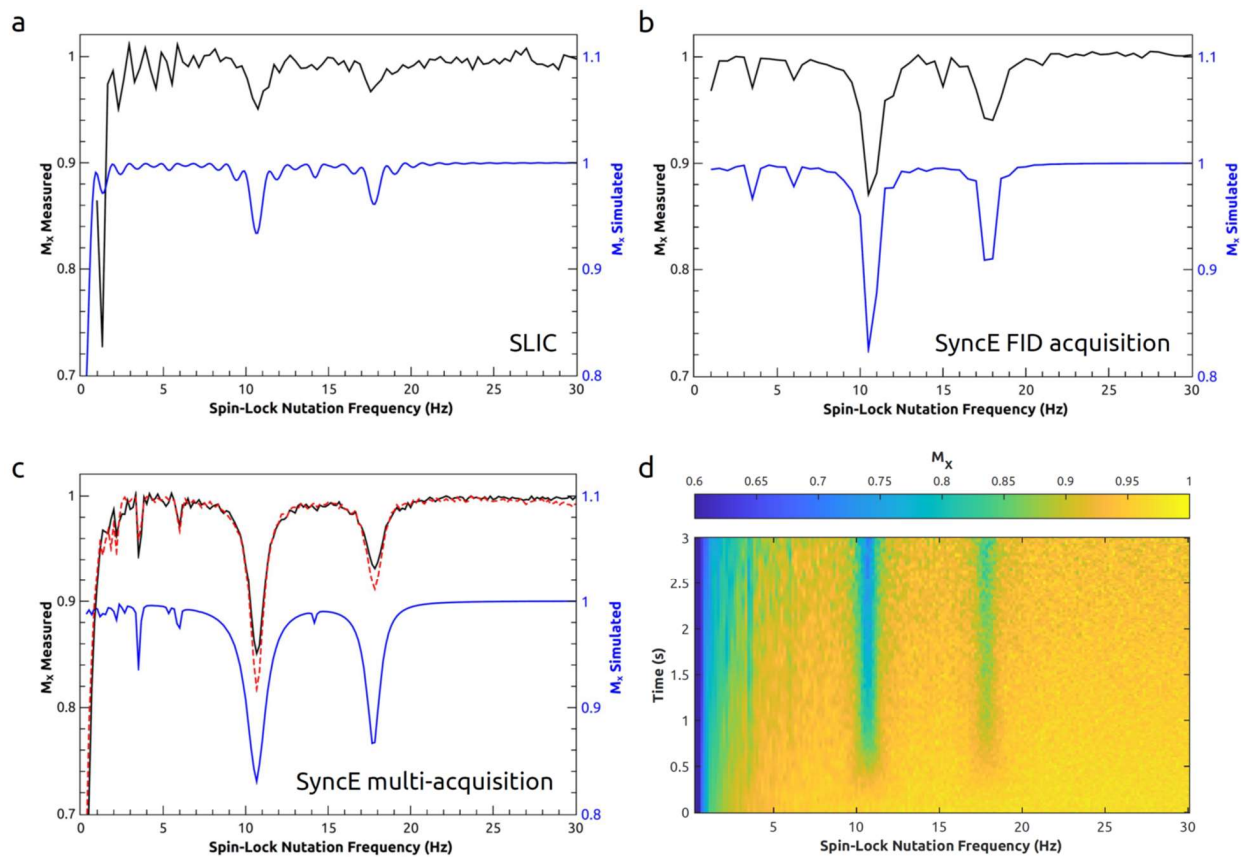

Figure 3. Measured (black) and simulated (blue) J-coupling spectra of neat ethyl acetate acquired with (a) SLIC, (b) synchronized echo (SyncE) using FID acquisitions, and (c) SyncE using the multi-acquisition scheme and averaging echoes. For the red dotted trace the hard pulses of the echo train were replaced with composite pulses. (d) 2D spectrum of ethyl acetate acquired with the multi-acquisition SyncE scheme. All data were acquired at 276 kHz (6.5 mT).

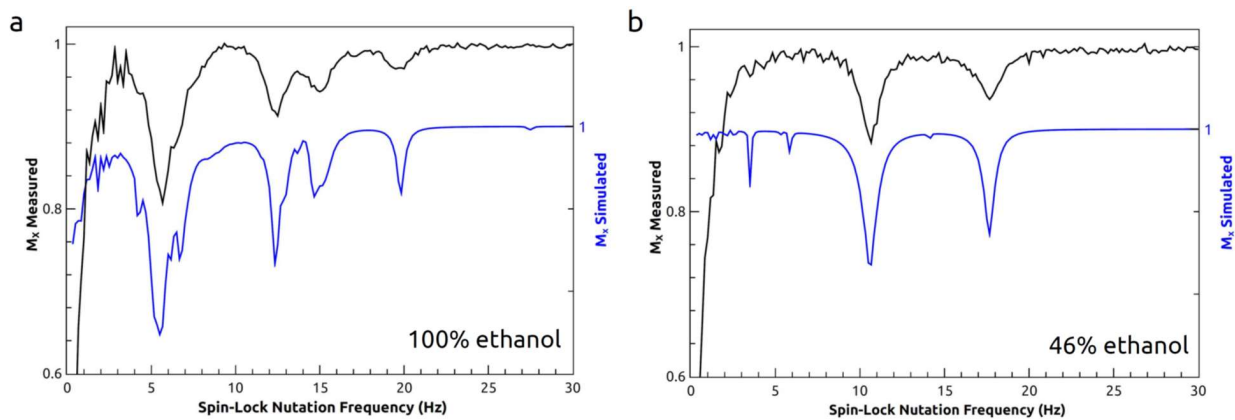

Figure 4. Measured (black) and simulated (blue) J-coupling spectra of (a) neat ethanol and (b) a mixture of 46% ethanol and 54% deionized water (%w/w). Spectra were acquired with the multi-acquisition SyncE scheme and all echoes were averaged to create the 1D projection. All data were acquired at 276 kHz (6.5 mT).

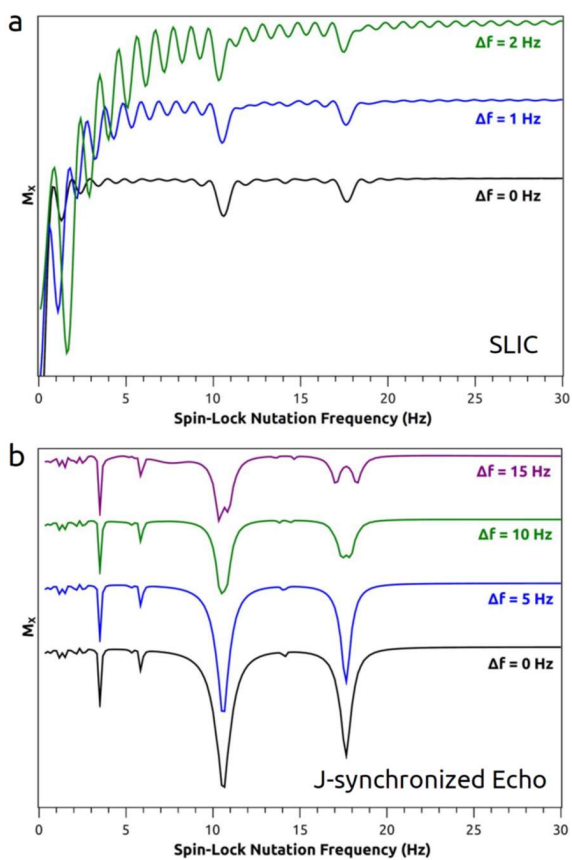

Figure 5: Simulated effects of $B_0$ imperfections on the simulated J-coupling spectra of hydrated ethanol. (a) SLIC with $B_0$ offsets between 0 and 2 Hz, using a 1 s evolution time. (b) Multi-acquisition synchronized echo with $B_0$ offsets between 0 and 15 Hz, using 2 ms refocusing pulses and a 3 s evolution time.

We also performed simulations to evaluate the sensitivity of the sequence to $B_0$ and $B_1$ errors. For SLIC, even a small offset in $B_0$ causes shifts in dip location as well as baseline oscillations, so $B_0$ must be carefully locked (Fig. 5a). The synchronized echo sequence is less sensitive to these errors, particularly if sufficiently short pulse lengths are used so that the flip angle maintains the necessary accuracy. Fig. 5b shows the effects of $B_0$ offsets when using 2 ms refocusing pulses. Much larger offsets are tolerated, before the dips eventually exhibit broadening and splitting.

$B_1$ offsets for spin-locking in SLIC lead to shifts in the dip locations (Fig. 6a). Since dip locations for synchronized echo are determined by delay times, which can be made highly accurate, it is more robust against such shifts. Instead, equivalent errors in the echo train pulse angles lead to broadening and splitting of the dips, similar to that observed for a $B_0$ offset. Figures 6b-c show that an accuracy of at least ±2° is needed for simple refocusing pulses, although this can be extended to ±20° or more with composite refocusing pulses. SyncE is also easier to calibrate than SLIC, as it requires only one $B_1$ calibration for the 180° pulse, rather than calibration for a series of spin-locking amplitudes with SLIC.

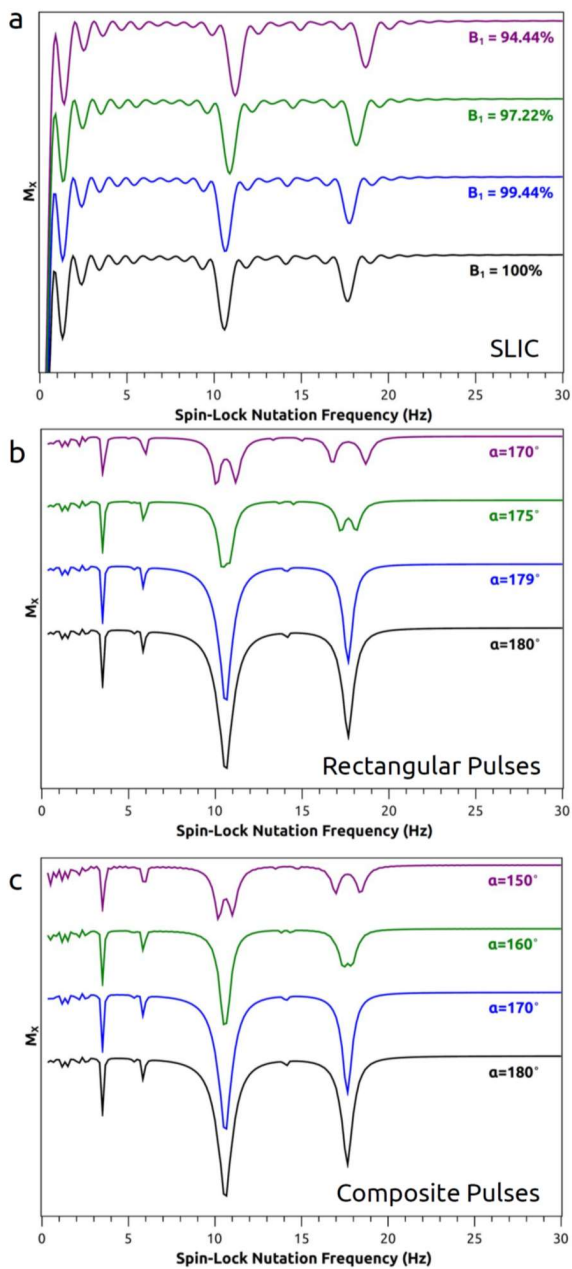

Figure 6: Simulated effects of $B_1$ imperfections on the J-coupling spectra of hydrated ethanol. (a) SLIC with $B_1$ errors between 0 and 5.56%. (b) Multi-acquisition synchronized echo using rectangular refocusing pulses with pulse angle errors between 0 and 10°, corresponding to $B_1$ offsets between 0 and 5.56%. (c) Multi-acquisition synchronized echo using $90_y$-$180_x$-$90_y$ composite refocusing pulses with pulse angle errors between 0 and 30°, corresponding to $B_1$ offsets between 0 and 16.68%.

While we chose CPMG phases for the synchronized echo pulse train (i.e. refocusing pulses shifted 90° from the initial excitation pulse), other phase choices for the refocusing pulses, such as y or alternating xy will also produce the same J-coupling spectra. In fact, using either y or xy eliminates the splitting caused by imperfect pulses. However, in those cases pulse imperfections instead produce oscillations in the baseline of the J-coupling spectrum as magnetization precesses around the y-axis of the Bloch sphere.

Finally, although our synchronized echo technique appears superficially similar to other CPMG pulse sequences, the strong-coupling regime leads to a very different response of the spin system at low field. For example, Freeman and Hill used a similar sequence to determine J-coupling at high field with high accuracy [26]. In that method, J-coupling leads to oscillations in echo intensity during the pulse train, which are converted to the spectral domain with a Fourier transform to find J-coupling values. These oscillations occur for any pulse spacing. In our technique, dynamics in echo intensity are instead driven by chemical shift differences, and the J-coupling values determine the pulse spacings at which these dynamics occur. This produces a spectrum of dips very similar to those acquired with quantum magnetometry using dynamical decoupling sequences [27-29], but again in those techniques dynamics are driven by dipolar or scalar couplings, and the frequencies at which dips occur correspond to the Larmor frequencies as spins precess about $B_0$.

**Conclusion**

Synchronized echo detection creates homonuclear J-coupling spectra with twice the signal strength of SLIC, and when combined with a multi-acquisition scheme increases the sensitivity by more than an order of magnitude. The reduced sensitivity to $B_0$ and $B_1$ errors helps make these synchronized echo approaches very robust even at the ultra-low (6.5 mT) field strengths used in the present work. Comparisons of dynamics between the synchronized echo and SLIC methods might also provide insights into different relaxation rates between dressed state populations versus coherences.

**Materials and Methods**

SLIC and synchronized echo measurements were performed on neat ethyl acetate, anhydrous ethanol, and a mixture of 46% ethanol with 54% deionized water (%w/w). All solvents were purchased from Sigma Aldrich.

Spectra at 276 kHz (6.5 mT) were measured in a custom-built high-homogeneity electromagnet-based MRI scanner with a Tecmag Redstone™ console described previously [30]. For the presently described work, a solenoidal sample coil was used, designed to hold 10 mm NMR tubes, and a $B_0$ field-frequency lock was used to maintain the resonance frequency within ±0.25 Hz. The scanner was shimmed to achieve a linewidth of deionized water of better than 0.5 Hz. RF pulses directly from the synthesizer were used, resulting in a 90° pulse length of 1 ms using about 4 µW. In most cases, a simple 2 ms 180° hard pulse was used for the spin echo pulse train, but a composite pulse consisting of the combination $90°_y - 180°_x - 90°_y$ was also evaluated, with the individual pulse lengths halved so that the total composite pulse was 2 ms long.

SLIC measurements were acquired using 1 s spin-locking time and nutation frequencies between 0.33 and 30 Hz with 0.33 Hz steps. Spin echo measurements were acquired with values of $\tau$ between 750 ms and 8.3 ms, giving a range of ~0 to 30 Hz for equivalent nutation frequency, $v_n$. The actual pulse delays were adjusted appropriately for the pulse width so that the time between the 180° pulse centers was $2\tau$. We used a total pulse train time $T = 1$ s with $n = 2$ to 60 loops or $T = 3$ s with $n = 2$ to 180 loops, giving resolutions of 0.5 Hz and 0.167 Hz, respectively. All pulses were performed on-resonance with the single $^1$H line of the conventional NMR spectrum. Pulses were calibrated with a Rabi experiment.

Echo acquisitions were 8 ms long centered between 180° pulses (16 points with 500 µs dwell time). FID acquisition was 5 s long with the same dwell time. The delay between measurements was at least 5 $T_1$.

To construct J-coupling spectra from both SLIC and synchronized echo FID data, each FID was converted to the spectral domain via the fast Fourier transform. The spectra were phased and integrated from -15 to 15 Hz. The integrals were divided by the maximal integrated signal from the whole set of spectra, and the result was plotted as a function of equivalent spin-lock nutation frequency to create a raw J-coupling spectrum. The $T_2$ background was then removed by dividing by a function

$$f(v_n) = A\left(1 - exp\left(-\frac{v_n}{B}\right)\right) + Cv_n + D \qquad (2)$$

where $v_n$ is the spin-lock nutation frequency, and $A, B, C,$ and $D$ are constants. Normally $A$ was between 0 and 1, $B$ was between 1 and 10, $C \approx 0$, and $D$ was between 0 and 1.5.

To construct a 2D spectrum from echo acquisitions, the magnitude of each echo was determined by averaging the middle 8 points of the magnitude spectrum for each echo. $T_2$ correction along the time axis was then performed by multiplying by the function $exp\left(\frac{t_k}{T_2}\right)$, where $t_k$ is the time at which echo $k$ is collected. Because the number of echoes acquired changes as a function of the sampled nutation frequency, a 2D plot of signal versus echo number and nutation frequency produces a triangular dataset. To visualize the data as a function of time instead of echo number, data were resampled along the time axis using spline interpolation.

From the 2D dataset, a 1D J-coupling spectrum was created by averaging across the time domain dimension, then dividing by the maximal signal from the whole set of averages, in the same manner as for FID acquisitions. The $T_2$ background was also removed in the same way.

Contrast-to-noise ratio (CNR) was calculated for ethyl acetate to compare the quality of the techniques. Contrast was calculated as the depth of each dip from the baseline $M_x = 1$, while noise was calculated as a standard deviation of measured values in the dip-free region between 22 and 30 Hz. To normalize for the different number of scans used, we calculated single scan CNR as $CNR = \frac{Contrast}{Noise \sqrt{N}}$, where $N$ is the number of scans performed. Noise was found to scale with $\sqrt{N}$ as expected.

SLIC and synchronized echo spectra were simulated using custom code written in MATLAB. The algorithm diagonalizes the Hamiltonian in the presence of a $B_1$ field, propagates the time dependent Schrodinger equation, and measures the remaining x-axis magnetization, $M_x$. As a check, some simulations were also performed with the Spin Dynamica package in Mathematica and the Spinach package in MATLAB [31-32]. A compiled version of our simulation software is also available online at https://github.com/ScalarMagnetics/SLIC-Simulator.